\documentclass[aps, prd, twocolumn, nofootinbib, showpacs]{revtex4}

\usepackage{amsfonts,amsmath,amsthm,amssymb,graphicx}

\usepackage{ulem}
\usepackage[usenames]{color}
\definecolor{DarkGreen}{rgb}{0.0,0.5,0.0}

\begin{document}

\preprint{1234}

\title{Inflationary potentials in DBI models}

\author{
        Dennis Bessada${}^{1,2}$,\footnote{
              {\tt dbessada@buffalo.edu},\\
              ${}^\dagger\,{}${\tt whkinney@buffalo.edu},\\${}^\ddagger\,{}${\tt ct38@buffalo.edu}
              }
              William~H.~Kinney${}^{1,\dagger}$, and Konstantinos~Tzirakis${}^{1,\ddagger}$}

     \affiliation{
              ${}^1$Dept. of Physics, University at Buffalo, the  State University of New York, Buffalo, NY 14260-1500, United States\\
              ${}^2$INPE - Instituto Nacional de Pesquisas Espaciais - Divis\~ao de Astrof\'isica, S\~ao Jos\'e dos Campos, 12227-010 SP, Brazil
             }

\date{\today}

\begin{abstract}

We study DBI inflation based upon a general model characterized by
a power-law flow parameter $\epsilon(\phi)\propto\phi^{\alpha}$
and speed of sound $c_s(\phi)\propto\phi^{\beta}$, where $\alpha$
and $\beta$ are constants. We show that in the slow-roll limit
this general model gives rise to distinct inflationary classes
according to the relation between $\alpha$ and $\beta$ and to the
time evolution of the inflaton field, each one corresponding to a
specific potential; in particular, we find that the well-known
canonical polynomial (large- and small-field), hybrid and
exponential potentials also arise in this non-canonical model. We
find that these non-canonical classes have the same physical
features as their canonical analogs, except for the fact that the
inflaton field evolves with varying speed of sound; also, we show
that a broad class of canonical and D-brane
inflation models are particular cases of this general
non-canonical model. Next, we compare the predictions of
large-field polynomial models with the current observational data,
showing that models with low speed of sound have red-tilted scalar
spectrum with low tensor-to-scalar ratio, in good agreement with
the observed values. These models also show a correlation between
large non-gaussianity with low tensor amplitudes, which is a
distinct signature of DBI inflation with large-field polynomial
potentials.

\end{abstract}


\maketitle

\section{\label{sec:intr}Introduction}

With the advent of the Five-year WMAP data \cite{komatsu2009} the inflationary paradigm \cite{starobinsky1979,guth1980,linde1981,albrecht1982}
(henceforth called {\it canonical inflation}) has been confirmed as the most successful candidate for explaining the physics of the very early
universe \cite{kinney2008}. The very rapid acceleration period generated by canonical inflation has solved some of the puzzles of the standard cosmological
model, such as
the horizon, flatness and entropy problems. However, existing models for inflation are phenomenological in character, and a fundamental
explanation of inflation is still missing. Also, the scalar field responsible for the inflationary expansion - the {\it inflaton} - is generically highly
fine-tuned.

Over the past few years developments in string theory have shed new light on these two problems of canonical inflation. String theory predicts a broad class of scalar fields associated with the compactification of extra dimensions and the configuration of
lower-dimensional branes moving in a higher-dimensional bulk space. This fact gave rise to some phenomenologically viable inflation models, such as the
KKLMMT scenario \cite{kachru2003}, Racetrack Inflation \cite{blancopillado2004}, Roulette Inflation
\cite{bond2006}, and the Dirac-Born-Infeld (DBI) scenario \cite{silverstein2003}. In particular, from a phenomenological point of view, the DBI scenario
has a very interesting and far-reaching feature: being a special case of a larger class of inflationary models with non-canonical Lagrangians
called {\it k-inflation} \cite{armendariz1999}, the DBI model possesses a varying speed of sound. This is a far-reaching feature because, in this case,
slow-roll can be achieved via a low sound speed instead of from dynamical friction due to expansion which, in turn, leads to
substantial non-Gaussianity \cite{alishahiha2004,chen2006,spalinski2007a,bean2007,loverde2007}. Also, DBI inflation admits several exact solutions to
the flow equations (first introduced in \cite{kinney2002} for the canonical case, and generalized in \cite{peiris2007} for DBI inflation), as
discussed in \cite{spalinski2007a,chimento2007,spalinski2007b,tzirakis2008a,tzirakis2009a}.

In this paper we derive a family of non-canonical models characterized by a power-law in the inflaton field $\phi$ for the speed of sound $c_{s}$ and the
flow parameter $\epsilon$.
We then derive
the forms of the associated inflationary potentials, and group them according to the classification scheme introduced in Ref. \cite{dodelson1997}, in order to have a better physical picture
of the solutions obtained. We later show that some particular cases possess spectral indices in agreement with the current observational
values, and establish the limits for the speed of sound in order that their tensor-to-scalar ratio correspond to the observed values. The outline
of the paper is as follows: in section \ref{sec:dbiinflation} we review the DBI inflation and the flow formalism for non-canonical inflation
with time-varying speed of sound. In section \ref{sec:powerspec} we review the tools needed to calculate the scalar and tensor spectral indices for
slow-roll. In section \ref{sec:powerlawmodel} we introduce the main features of our model and derive general solutions, postponing
the discussion of their physical properties to section \ref{sec:inflpotentials}, where we extend the concept of large-field, small-field, hybrid and
exponential potentials of canonical inflation to the non-canonical case. In section \ref{sec:application} we compare the observational predictions of a set of
non-canonical large-field models with WMAP5 data, and show that there is a generic correlation between small sound speed (and therefore significant non-Gaussianity) and a low tensor/scalar ratio. We present conclusions in \ref{sec:conclusions}.

\section{\label{sec:dbiinflation}DBI Inflation - An Overview}

In warped D-brane inflation (see \cite{mcallister2007} and \cite{cline2006} for a review), inflation is regarded as the motion of a D3-brane in a
six-dimensional ``throat" characterized by the metric \cite{klebanov2000}
\begin{equation}
\label{eq:DBImetric} ds^2_{10} = h^2\left(r\right) ds^2_4 +
h^{-2}\left(r\right) \left(d r^2 + r^2 ds^2_{X_5}\right),
\end{equation}
where $h$ is the warp factor, $X_5$ is a Sasaki-Einstein five-manifold which forms the base of the cone, and $r$ is the radial coordinate along the throat. In this case, the
inflaton field $\phi$ is identified with $r$ as $\phi = \sqrt{T_3} r$, where $T_3$ is the brane tension. The dynamics of the D3-brane in the warped
background (\ref{eq:DBImetric}) is then dictated by the DBI Lagrangian
\begin{equation}
\label{DBIlagrangian} {\cal L} = - f^{-1}\left(\phi\right) \sqrt{1
- 2 f\left(\phi\right) X } - f^{-1}\left(\phi\right) -
V\left(\phi\right),
\end{equation}
where $g_{\mu \nu}$ is the background spacetime metric, $f^{-1}(\phi)=T_3h(\phi)^4$
is the inverse brane tension, $V(\phi)$ is an arbitrary potential,
and $X=(1/2)g^{\mu \nu}\partial_{\mu}\phi
\partial_{\nu}\phi$ is the kinetic term. We assume that the
background cosmological model is described by the flat
Friedmann-Robertson-Walker (FRW) metric, $g_{\mu \nu}={\rm
diag}\left\{1,-a^2(t),-a^2(t),-a^2(t)\right\}$.
DBI inflation is a
special case of {\it k-inflation}, characterized by a varying
speed of sound $c_s$, whose expression is given
by
\begin{equation}
\label{defspeedofsound} c_s^{2} = \left(1 + 2X\frac{{\cal
L}_{,XX}}{{\cal L}_{,X}}\right)^{-1},
\end{equation}
where the subscript ``${,X}$" indicates a derivative with respect to the kinetic
term. It is straightforward to see that in the model
(\ref{DBIlagrangian}) the speed of sound is given by
\begin{equation}
\label{DBIspeedofsound} c_s(\phi) = \sqrt{1 - 2f(\phi)X}.
\end{equation}
In those models it is convenient to
introduce an analog of the Lorentz factor, related to
$c_s(\phi)$ by:
\begin{equation}
\label{gammafac} \gamma(\phi) = \frac{1}{c_s(\phi)}.
\end{equation}

We next introduce the
generalization of the inflationary flow hierarchy for k-inflation models
\cite{peiris2007}. The two fundamental flow parameters are
\begin{subequations}
\begin{eqnarray}
\label{defepsilon1} \epsilon\left(\phi\right) &=& \frac{2
M_P^2}{\gamma\left(\phi\right)}
\left(\frac{H'\left(\phi\right)}{H\left(\phi\right)}\right)^2,\\
\label{defs1} s\left(\phi\right) &=& \frac{2
M_P^2}{\gamma\left(\phi\right)}
\frac{H'\left(\phi\right)}{H\left(\phi\right)}
\frac{\gamma'\left(\phi\right)}{\gamma\left(\phi\right)},
\end{eqnarray}
\end{subequations}
where the prime
indicates a derivative with respect to $\phi$. An infinite hierarchy of additional flow parameters can be generated by differentiation, and are defined as follows:
\begin{eqnarray}
\label{eq:flowparams}
\eta\left(\phi\right) &=& \frac{2
M_P^2}{\gamma\left(\phi\right)}
\frac{H''\left(\phi\right)}{H\left(\phi\right)},\cr &\vdots& \cr {}^\ell \lambda\left(\phi\right) &=& \left(\frac{2
M_P^2}{\gamma\left(\phi\right)}\right)^{\ell}
\left(\frac{H'\left(\phi\right)}{H\left(\phi\right)}\right)^{\ell
- 1} \frac{1}{H\left(\phi\right)} \frac{d^{\ell + 1}
H\left(\phi\right)}{d \phi^{\ell + 1}},\cr\cr
\rho\left(\phi\right) &=& \frac{2
M_P^2}{\gamma\left(\phi\right)}
\frac{\gamma''\left(\phi\right)}{\gamma\left(\phi\right)},\cr &\vdots& \cr
{}^\ell
\alpha\left(\phi\right) &=& \left(\frac{2
M_P^2}{\gamma\left(\phi\right)}\right)^{\ell}
\left(\frac{H'\left(\phi\right)}{H\left(\phi\right)}\right)^{\ell
- 1} \frac{1}{\gamma\left(\phi\right)} \frac{d^{\ell + 1}
\gamma\left(\phi\right)}{d \phi^{\ell + 1}},
\end{eqnarray}
where $\ell = 2,\ldots,\infty$ is an integer index. It is convenient to
express the flow parameters (\ref{eq:flowparams}) in terms of the
number of e-folds before the end of inflation, which is defined in
the following way:
\begin{equation}
\label{eq:numefolds} N = - \int{H}{dt} = \frac{1}{\sqrt{2 M_P^2}}
\int_{\phi_e}^{\phi}{\sqrt{\frac{\gamma\left(\phi\right)}{\epsilon\left(\phi\right)}}
d\phi},
\end{equation}
where $\phi_e$ indicates the value of the inflaton field at the
end of inflation. The definition above indicates that the number
of e-folds increases as we go backward in time, so that it is zero
at the end of inflation. We can also determine the scale factor, $a$, in terms of $N$ in a straightforward way: since $da/a=dtH=-dN$, we simply have
\begin{equation}
\label{aefolds}
a(N)=a_e e^{-N}.
\end{equation}

In terms of $N$, it is easy to see that the key flow
parameters $\epsilon$ and $s$, as defined in (\ref{defepsilon1}) and (\ref{defs1}),
assume the following equivalent form
\begin{subequations}
\begin{eqnarray}
\label{defepsilon2} \epsilon &=& \frac{1}{H}\frac{d H}{d N},\\
\label{defs2} s &=& \frac{1}{\gamma}\frac{d \gamma}{d N},
\end{eqnarray}
\end{subequations}
so that we may derive a set of first-order differential
equations
\begin{eqnarray}
\label{eq:flowequations} \frac{d \epsilon}{d N} &=&
\epsilon\left(2 \eta - 2 \epsilon - s\right),\cr \frac{d \eta}{d
N} &=& -\eta\left(\epsilon + s\right) + {}^2 \lambda,\cr &\vdots& \cr \frac{d
{}^\ell \lambda}{d N} &=& -{}^\ell \lambda \left[\ell \left(s +
\epsilon\right) - \left(\ell - 1\right) \eta\right] + {}^{\ell +
1}\lambda, \cr \frac{d s}{d N} &=& -s \left(2 s + \epsilon -
\eta\right) + \epsilon \rho,\cr \frac{d \rho}{d N} &=& -2 \rho s +
{}^2 \alpha,\cr &\vdots& \cr \frac{d {}^\ell \alpha}{d N} &=& -{}^\ell \alpha
\left[\left(\ell + 1\right) s + \left(\ell - 1\right)
(\epsilon-\eta)\right] + {}^{\ell + 1} \alpha.
\end{eqnarray}
The dynamics of the inflaton field can be completely described by this hierarchy of equations, which are equivalent to
the Hamilton-Jacobi equations \cite{spalinski2007c}
\begin{subequations}
\begin{eqnarray}
\label{eq:Hamjacobi1} \dot
\phi&=&-\frac{2M_P^2}{\gamma(\phi)}H'(\phi),\\
\label{eq:Hamjacobi2}3M_P^2H^2(\phi)&-&V(\phi)=\frac{\gamma(\phi)-1}{f(\phi)},
\end{eqnarray}
\end{subequations}
where $M_P=1/\sqrt{8\pi G}$ is the reduced Planck mass, and
\begin{equation}
\label{eq:Hamjacobi 2} \gamma(\phi)=\sqrt{1+4M_P^4f(\phi)\left[H'(\phi)\right]^2}.
\end{equation}
In terms of the flow parameter $\epsilon$, the potential, $V(\phi)$, and the inverse brane tension, $f(\phi)$, can be written as
\begin{equation}
\label{Vgen}
V\left(\phi\right) = 3 M_P^2 H^2 \left(1 - \frac{2\epsilon}{3}
\frac{\gamma}{\gamma+1}\right),
\end{equation}
and
\begin{equation}
\label{fgen} f\left(\phi\right) = \frac{1}{2 M_{P}^2
H^2\epsilon}\left(\frac{\gamma^2 - 1}{\gamma}\right),
\end{equation}
respectively. Our strategy in this paper is to generate inflationary solutions by {\it ansatz} for the flow parameters $\epsilon$ and $s$, for which the forms of the functions $V\left(\phi\right)$ and $f\left(\phi\right)$ are determined.

In the next section, we briefly review the generation of perturbations in non-canonical inflation models.

\section{\label{sec:powerspec} Power Spectra}

Scalar and tensor perturbations in k-inflation were first studied in Ref. \cite{garriga1999}, where it was shown that the quantum
mode function $u_k$ obeys the equation of motion
\begin{equation}
\label{modeeq} u_k'' + \left(c_s^2 k^2 - \frac{z''}{z}\right) u_k =
0,
\end{equation}
where $k$ is the comoving wave number, a prime denotes a derivative with respect to conformal time
$d\tau \equiv dt / a$, and $z$ is a variable given by
\begin{equation}
z = \frac{a \sqrt{\rho + p}}{c_s H} = \frac{a \gamma^{3/2}
\dot\phi}{H} = - M_p a \gamma \sqrt{2 \epsilon}.
\end{equation}
It is convenient to change the conformal time, $\tau$, to
\begin{equation}
y \equiv \frac{k}{\gamma a H},
\end{equation}
so that
\begin{equation}\label{dery}
\frac{d^2}{d\tau^2} = a^2 H^2 \left[\left(1 - \epsilon -
s\right)^2 y^2 \frac{d^2}{dy^2} +
G\left(\epsilon,\eta,s,\rho\right) y \frac{d}{d y}\right],
\end{equation}
where
\begin{eqnarray}
\label{defG} G &=& -s + 3 \epsilon s - 2
\epsilon \eta - \eta s + 2\epsilon^2 + 3 s^2 - \epsilon \rho.
\end{eqnarray}
We can also show that
\begin{equation}
\label{zF}
\frac{z''}{z} = a^2 H^2
F\left(\epsilon,\eta,s,\rho,{}^2\lambda\right),
\end{equation}
where \cite{shandera2006}
\begin{eqnarray}
\label{defF} F = &&2 + 2 \epsilon - 3 \eta - \frac{3}{2} s - 4
\epsilon \eta + \frac{1}{2} \eta s  + 2 \epsilon^2\cr
 &&+ \eta^2 - \frac{3}{4} s^2 + {}^2\lambda + \frac{1}{2}\epsilon\rho.
\end{eqnarray}
Substituting (\ref{dery}) and (\ref{zF}) into (\ref{modeeq}), we find
\begin{equation}
\label{exactmode} \left(1 - \epsilon -s\right)^2 y^2 \frac{d^2
u_k}{dy^2} + G y \frac{d u_k}{d y} + \left[y^2 - F\right] u_k = 0,
\end{equation}
which is an exact equation, without any assumption of slow-roll.

We now proceed to solve equation (\ref{exactmode}) to first-order in slow-roll. In this case, this equation takes the form
\begin{eqnarray}
\label{modeeqsr} (1&- 2\epsilon &-2s)y^2 \frac{d^2
u_k}{dy^2}-s y \frac{d u_k}{d y}\cr&& + \left[y^2 -
2(1+\epsilon-\frac{3}{2}\eta-\frac{3}{4}s)\right] u_k = 0,
\end{eqnarray}
whose solution is \cite{tzirakis2008a}
\begin{equation}
\label{uksolsr}
u_{k}(y)=\frac{1}{2}\sqrt{\frac{\pi}{c_{s}k}}\sqrt{\frac{y}{1-\epsilon-s}}H_{\nu}\left(\frac{y}{1-\epsilon-s}\right),
\end{equation}
where $H_\nu$ is a Hankel function, and
\begin{equation}
\label{defnu} \nu=\frac{3}{2}+2\epsilon-\eta+s.
\end{equation}

The power spectra for scalar and tensor perturbations are given respectively by
\begin{eqnarray}
\label{powerspectra}
P_{{\cal R}}&=&\frac{1}{8\pi^{2}}\left.\frac{H^2}{M_P^2 c_s\epsilon}\right|_{c_sk=aH},\cr
P_{T}&=&\frac{2}{\pi^{2}}\left.\frac{H^2}{M_P^2 }\right|_{k=aH},
\end{eqnarray}
so that, for the solution (\ref{uksolsr}),
\begin{eqnarray}
\label{scpowerspecsr}
P_{\mathcal{R}}^{1/2}&=&[(1-\epsilon-s)\cr&&+(2-\ln2-\gamma)(2\epsilon-\eta+s)]\left.\frac{H^2}{2\pi
\dot\phi}\right|_{k=\gamma a H}
\end{eqnarray}
where $\gamma \approx 0.577$ is Euler's constant. From the definitions of the scalar and tensor spectral indices,
\begin{eqnarray}
\label{specindices} n_{s}-1 &\equiv&
\frac{d(\ln P_{\mathcal{R}})}{d(\ln k)},\nonumber \\
n_{T} &\equiv&
\frac{d(\ln P_{T})}{d(\ln k)},
\end{eqnarray}
we find
\begin{eqnarray}
\label{scalarspecindex} n_{s}-1 &=& -4\epsilon + 2\eta -2s, \nonumber \\
n_{T} &=& -2\epsilon,
\end{eqnarray}
valid to first-order in the slow-roll limit. The tensor-to-scalar ratio is given by \cite{Powell:2008bi}
\begin{equation}
\label{stratiod} r =16\epsilon c_s^{\frac{1+\epsilon}{1-\epsilon}},
\end{equation}
which, to first-order in slow-roll, gives the well-known result
\begin{equation}
\label{stratio} r =16c_s\epsilon,
\end{equation}
Since in this paper we will work only to the first-order in slow-roll approximation, we make use of expression (\ref{stratio}) to compute the
tensor/scalar ratio. It is immediately clear from this expression that small $c_s$ has the generic effect of suppressing the tensor/scalar ratio.

\section{\label{sec:powerlawmodel}The Model}

\subsection{\label{subsec:gensetting}The General Setting}

The usual approach to the construction of a model of inflation normally starts with a choice of the inflaton potential, $V(\phi)$;
then, all the flow parameters are derived, and the dynamical analysis is performed. In this work we adopt the reverse procedure: we first look for the
solutions to the differential equation satisfied by the Hubble parameter $H(\phi)$,
\begin{eqnarray}
\label{eqhubble} \frac{H'(\phi)}{H(\phi)}&=&\pm \sqrt{\frac{\epsilon(\phi)\gamma(\phi)}{2M_P^2}},
\end{eqnarray}
and only afterwards derive the form of the potential. Equation (\ref{eqhubble}) can be easily derived from the definition of the
flow parameter $\epsilon$, given by (\ref{defepsilon1}), and the sign ambiguity indicates in which direction the field is rolling. Notice that in
order to solve equation (\ref{eqhubble})
we must know the form of the functions $\epsilon(\phi)$ and $\gamma(\phi)$; we choose them to be power-law functions of the inflaton field,
\begin{subequations}
\begin{eqnarray}
 \epsilon(\phi)&=&\left(\frac{\phi}{\phi_e}\right)^\alpha,\label{powerepsilon}\\\gamma(\phi)&=&\gamma_e\left(\frac{\phi}{\phi_e}\right)^\beta,
 \label{powercs}
\end{eqnarray}
\end{subequations}
where $\gamma_e$ is the value of the Lorentz factor at the end of inflation\footnote{Henceforth all the variables with a subscript {\it{e}} are
evaluated at the end of inflation.}, and $\alpha$ and $\beta$ are constants. Another case worth studying appears when $\epsilon$ is {\it{constant}},
so that
\begin{eqnarray}
 \epsilon(\phi)&=&\epsilon = {\rm const.},\label{powerepsilonrep}
\end{eqnarray}
with the same parametrization for $\gamma$. We have kept $\phi_e$ for the following reason: in the $IR$ DBI model the inflaton field rolls down from
the tip of the throat toward the bulk of
the manifold with increasing speed of sound; then, when the field enters the bulk $c_s$ becomes equal to $1$, and then inflation ``ends". In the $UV$
case the behavior is the opposite, that is, the field evolves away from the bulk and reaches the tip when $c_s=1$.
To reproduce both cases we could have set $\gamma_e=1$ from the onset, so that $c_s(\phi_e)=1$, as required; also, $c_s(\phi)=1$ in the canonical limit,
that is, when $\beta=0$. However, by taking $\gamma_e$ arbitrary, we also reproduce the non-canonical models with constant speed of sound
introduced by Spalinski \cite{spalinski2007b}. It is clear that in the latter case $\gamma_e$ does not refer necessarily to the end of inflation,
so that if we take $\gamma_e\gg 1$ it does not mean a superluminal propagation (as would be the case if $\beta\neq0$). Bearing this distinction in mind
we can use the same notation unambiguously.

When $\alpha\neq0$, substituting
(\ref{powerepsilon}) and (\ref{powercs}) into (\ref{eqhubble}), we see that the Hubble parameter takes the form
\begin{eqnarray}
\label{eqhubble1} H(\phi)&=&H_e\exp \left[\sigma\sqrt{\frac{\gamma_e}{2M_P^{2}\phi_e^{\alpha+\beta}}}I(\phi)\right],
\end{eqnarray}
where we have defined
\begin{equation}
\label{eqhubble2} I(\phi)\equiv\int^{\phi}_{\phi_e}d\phi' {\phi'}^{(\alpha+\beta)/2},
\end{equation}
and $\sigma$ accounts for the sign ambiguity appearing in (\ref{eqhubble}). When $\epsilon$ is constant, the solution to
(\ref{eqhubble}) reads
\begin{eqnarray}
\label{hubblerep0} H(\phi)&=&H_e\exp \left[\sigma\sqrt{\frac{\epsilon\gamma_e}{2M_P^{\beta+2}}}I(\phi)\right],
\end{eqnarray}
where the integral $I(\phi)$ is the same as in
(\ref{eqhubble2}).

It is clear that the integral (\ref{eqhubble2}) admits two distinct solutions: a logarithmic one when $\alpha+\beta=-2$, and power-law
for $\alpha+\beta\neq -2$. These two solutions will give rise to different classes of inflationary potentials, which we shall address in the
next subsections.

To conclude this section we derive the general formula for the number of e-folds. For $\alpha\neq0$ this expression can be determined by equations
(\ref{eq:numefolds}),
(\ref{powerepsilon}), and (\ref{powercs}), so that
\begin{equation}
\label{efolds1} N(\phi)=\sigma\sqrt{\frac{\gamma_e
\phi_e^{\alpha-\beta}}{2M_P^2}}J(\phi),
\end{equation}
where
\begin{equation}
\label{efolds2} J(\phi)=\int^{\phi}_{\phi_e}d\phi'
{\phi'}^{(\beta-\alpha)/2};
\end{equation}
if $\alpha=0$ we must use the parametrization (\ref{powerepsilonrep}), so that expression (\ref{efolds1}) changes to
\begin{equation}
\label{efolds3}
N(\phi)=\sigma\sqrt{\frac{\gamma_e
}{2M_P^2\epsilon\phi_e^{\beta}}}\tilde{J}(\phi),
\end{equation}
where
\begin{equation}
\label{efolds4}
 \tilde{J}(\phi)=\int^{\phi}_{\phi_e}d\phi'
{\phi'}^{\beta/2}.
\end{equation}

In the next section, we discuss particular cases of this general class of solutions.

\subsection{\label{subsec:sol1}The Solutions {\boldmath{$\alpha+\beta=-2$}}.}

For this class of solutions, the parameters $\alpha$ and $\beta$ are related by
\begin{equation}
\label{sol1alpbet}
\alpha=-2-\beta;
\end{equation}
notice that the case $\alpha=0$ implies $\beta=-2$, which corresponds to the case where the flow parameters $\epsilon$ and $s$ are constant
\cite{tzirakis2008a}. Evaluating the integral (\ref{eqhubble2}) and substituting
it into (\ref{eqhubble1}), we find
\begin{equation}
\label{sol1hubpar} H(\phi)=H_e\left(\frac{\phi}{\phi_e}\right)^{p/2},
\end{equation}
where the exponent $p$ is determined by
\begin{equation}
\label{sol1defp}
p=\sigma \phi_e\sqrt{\frac{2\gamma_e}{M_P^{2}}}.
\end{equation}
Let us now analyze the sign ambiguity appearing in (\ref{sol1defp}). We first write the expression (\ref{sol1hubpar}) as
\begin{equation}
\label{sol1defp1} p=2\frac{\ln\left(H(\phi)/H_e\right)}{\ln\left(\phi/\phi_e\right)};
\end{equation}
then, for $\alpha<0$, we have, from (\ref{powerepsilon}), in the slow-roll limit,
\begin{equation}
\label{sol1epslf}
\epsilon(\phi)=\left(\frac{\phi}{\phi_e}\right)^{-|\alpha|}\ll 1,
\end{equation}
which implies $\phi\gg\phi_e$ and $\dot{\phi}<0$ for $\phi>0$ ({\it{large-field limit}}), so that $\ln\left(\phi/\phi_e\right)>0$.
Since the weak-energy condition implies that $\dot H<0$, we have
$H(\phi)\geq H_e$, and then $\ln\left(H(\phi)/H_e\right)>0$. Therefore, from (\ref{sol1defp1}), $\alpha<0$ implies $p>0$. From definition
(\ref{sol1defp}) we see that $p>0$ implies $\sigma=-1$ if $\phi<0$, and $\sigma=+1$ if $\phi>0$.

Conversely, if
$\alpha>0$, we have, from (\ref{powerepsilon}), in the slow-roll limit,
\begin{equation}
\label{sol1epssf}
\epsilon(\phi)=\left(\frac{\phi}{\phi_e}\right)^{+|\alpha|}\ll 1,
\end{equation}
which implies $\phi\ll\phi_e$ and $\dot{\phi}>0$ for $\phi>0$ ({\it{small-field limit}}), so that $\ln\left(\phi/\phi_e\right)<0$.
Again, $\ln\left(H(\phi)/H_e\right)>0$, so that, from (\ref{sol1defp1}), $\alpha>0$ implies $p<0$. From definition
(\ref{sol1defp}) we see that $p<0$ implies $\sigma=-1$ if $\phi>0$, and $\sigma=+1$ if $\phi<0$.

Therefore, what really distinguishes the models is the sign of the exponent of $\alpha$; the sign of $\sigma$ simply dictates which direction the
field is rolling in, exactly as in canonical inflation. We are left with two distinct models, whose properties are summarized in Table \ref{t1}.
\begin{table}[htbp]
\caption{The sign rule for models with $\alpha+\beta=-2$ and $\phi>0$.}\label{t1}
\begin{tabular}{|c||c|c|c|c|} \hline \hline
Model 1. &$p>0$, $\alpha<0$ & $\sigma=+1$ & $\phi>0$  & $\dot\phi<0$\\ \hline
Model 2. &$p<0$, $\alpha>0$ & $\sigma=+1$ & $\phi>0$  & $\dot\phi<0$\\ \hline
\end{tabular}
\end{table}
(For $\phi<0$ the sign rule is easily obtained by flipping all the signs of the quantities present in Table \ref{t1}.)

\subsection{\label{subsec:sol2}The Solutions {\boldmath{$\alpha+\beta\neq-2$}} }

Let us first consider the case $\alpha\neq0$. The solution to the integral (\ref{eqhubble2}) is
\begin{equation}
\label{sol2hubpar}
H(\phi)={\tilde{H}}_e\exp\left[\frac{\sigma K\phi_e}{\alpha+\beta+2}{\bar{\phi}}^{(\alpha+\beta+2)/2}
\right],
\end{equation}
where we have defined
\begin{equation}
\label{sol2hubpare}
{\tilde{H}}_e=H_e\exp\left[-\frac{\sigma\sqrt{2\gamma_e}\phi_e}{M_P(\alpha+\beta+2)}
\right],
\end{equation}
\begin{equation}
\label{sol2defKpb}
K=\sqrt{\frac{2\gamma_e}{M_P^2}}, ~~~~~{\bar{\phi}}=\frac{\phi}{\phi_e}.
\end{equation}
In order to fix the sign ambiguity let us rewrite expression (\ref{sol2hubpar}) as
\begin{equation}
\label{sol2lnhubpar1}
d\ln H=\frac{2\sigma K\phi_e}{\alpha+\beta+2}d{\bar{\phi}}^{(\alpha+\beta+2)/2};
\end{equation}
then, as we have seen in (\ref{sol1epslf}), the condition $\alpha<0$ corresponds to the large-field limit, $\phi\gg\phi_e$; so,
if $\alpha+\beta+2<0$ and $\phi>0$, we have $d{\bar{\phi}}^{(\alpha+\beta+2)/2}>0$, so that from (\ref{sol2lnhubpar1}),
\begin{eqnarray}
\label{sol2lnhubpar}
&\underbrace{d\ln H}_{<0}&=\frac{2\sigma K\phi_e}{\alpha+\beta+2}\underbrace{d{\bar{\phi}}^{(\alpha+\beta+2)/2}}_{>0}\nonumber \\
&\Longrightarrow& -\frac{\sigma K|\phi_e|}{|\alpha+\beta+2|}<0,
\end{eqnarray}
implying that $\sigma=+1$. Applying the same analysis for $\alpha+\beta+2<0$ and $\phi<0$, we find $\sigma=-1$. For $\alpha+\beta+2>0$ we find
$\sigma=+1$ for $\phi>0$, and $\sigma=-1$ for $\phi<0$. The results in the large-field limit are summarized in Table
\ref{t2}. The same reasoning also applies for $\alpha>0$, that is, the small-field limit, given by (\ref{sol1epssf});
the results are summarized in Table \ref{t3}, where we have four distinct models:
\begin{table}[htbp]
\caption{The sign rule for models with $\alpha+\beta\neq -2$, $\alpha<0$, and $\phi>0$.}\label{t2}
\begin{tabular}{|c||c|c|c|} \hline \hline
Model 3. &$\alpha+\beta+2>0$ & $\sigma=+1$ & $\dot\phi<0$\\ \hline
Model 4. &$\alpha+\beta+2<0$ & $\sigma=+1$ & $\dot\phi<0$\\ \hline
\end{tabular}
\end{table}
\begin{table}[htbp]
\caption{The sign rule for models with $\alpha+\beta\neq0$, $\alpha>0$ and $\phi>0$.}\label{t3}
\begin{tabular}{|c||c|c|c|} \hline \hline
Model 5. &$\alpha+\beta+2>0$ & $\sigma=-1$ & $\dot\phi>0$\\ \hline
Model 6. & $\alpha+\beta+2>0$ & $\sigma=+1$ & $\dot\phi<0$\\ \hline
Model 7. &$\alpha+\beta+2<0$ & $\sigma=-1$ & $\dot\phi>0$\\ \hline
Model 8. & $\alpha+\beta+2<0$ & $\sigma=+1$ & $\dot\phi<0$\\ \hline
\end{tabular}
\end{table}
As in the previous case, the model $\phi<0$
is easily obtained by flipping all the signs of the quantities present in Tables \ref{t2} and \ref{t3}.

In the case $\epsilon=const.$, $\beta\neq-2$, the solution to the integral (\ref{eqhubble2}) is given by
\begin{equation}
\label{sol2hubparalp0}
H(\phi)={\tilde{H}}_e\exp\left[\sigma\sqrt{\frac{2\epsilon\gamma_e}{M_P^2\phi_e^{\beta}}}\frac{\phi^{(\beta+2)/2}}
{\beta+2}\right],
\end{equation}
where
\begin{equation}
\label{sol2hubparealp0}
{\tilde{H}}_e=H_e\exp\left[-\sigma\sqrt{\frac{2\epsilon\gamma_e}{M_P^2\phi_e^{\beta}}}\frac{\phi_e^{(\beta+2)/2}}
{\alpha+\beta+2}\right].
\end{equation}
It is straightforward to see that the same sign rules shown in Tables \ref{t2} and \ref{t3} also apply to this case.

\section{\label{sec:inflpotentials}Classes of inflationary potentials in DBI inflation}

In this section we proceed to analyze the solutions obtained
in the last section. The key ingredient, in order to
understand the physics associated with these classes of solutions,
is the study of the form of the inflationary potential, which is
obtained from the expressions (\ref{Vgen}), (\ref{powercs}), and
the corresponding expression for the Hubble parameter, given by
either (\ref{sol1hubpar}) or (\ref{sol2hubpar}). Once we have the form of such non-canonical
potentials, we can compare these expressions with the usual
canonical inflationary potentials,
grouped into four
general classes: the {\it{large-}} and {\it{small-field polynomial}}, {\it{hybrid}},
and {\it{exponential}} models \cite{dodelson1997}. To do so, we must make some choices for the exponents $\alpha$ and $\beta$ first, and then on the
corresponding dynamics of the field; hence, as we have seen in the tables \ref{t1}, \ref{t2} and \ref{t3}, we have eight distinct models altogether.
Our main aim is to reproduce the classes of the inflationary potentials discussed in \cite{dodelson1997}; then, in doing so,
we leave out some interesting solutions, but our emphasis here is on understanding the physics of the non-canonical models, which can be achieved
through a close comparison with the well-established potentials found in the literature.

\subsection{\label{subsec:largefield}Large-field polynomial potentials}

In these models, the inflaton field is displaced far from its minimum to a value $\phi\sim \mu$, and then rolls down toward its minimum
at the origin on a potential $V(\phi)\propto \phi^p$. A quick look at expressions (\ref{Vgen}) and (\ref{sol1hubpar}) suggests that the
model $1$ in table \ref{t1}, characterized by
$p>0$ and $\alpha<0$, is its non-canonical counterpart, for its potential also goes like $\phi^p$. To check this, let us first analyze the behavior
of the speed of sound (\ref{powercs}).
Since the equality (\ref{sol1alpbet}) implies $\beta > -2$, we see from
(\ref{powercs}) that $\beta > 0$ corresponds to
\begin{equation}
\label{lfieldgamm2}
\gamma(\phi)=\gamma_e \left(\frac{\phi}{\phi_e}\right)^{|\beta|} \Longrightarrow \gamma \rightarrow \infty ~~~\mathrm{as}~~~\phi\rightarrow \infty,
\end{equation}
or, in terms of the speed of sound,
\begin{eqnarray}
\label{lfieldcs3}
c_s &\rightarrow& 0 ~~~\mathrm{as}~~~\phi\rightarrow \infty;
\end{eqnarray}
since in the large-field limit the field strength is very large at early times, we conclude from (\ref{lfieldcs3}) that
the speed of sound starts off with a {\it subluminal} value. Also, from (\ref{lfieldgamm2}), we see that $\gamma/(\gamma+1)\rightarrow 1$; then,
using this fact and plugging (\ref{sol1hubpar}) into (\ref{Vgen}),
we find
\begin{eqnarray}
\label{lfieldpot2} V(\phi)&\sim& 3M_P^2H_e^2\left(\frac{\phi}{\phi_e}\right)^{p},
\end{eqnarray}
which has exactly the same form of a canonical large-field
potential. The non-canonical potential (\ref{lfieldpot2}) shows
that the inflaton field starts evolving from a value $\phi\sim
\mu$ with a very low speed of sound, and then rolls down
toward its minimum at origin.
Once there, the speed of sound becomes unity as well as the flow
parameter $\epsilon$ and then inflation ends. The potential
evaluated at $\mu$ corresponds to the vacuum energy density,
\begin{equation}
\label{lfieldhpot} V(\mu)=\Lambda^4\Longrightarrow \Lambda^4\sim
3M_P^2H_e^2,
\end{equation}
so that in terms of these two quantities, the Hubble parameter
(\ref{sol1hubpar}) and the inflationary potential
(\ref{lfieldpot2}) assume the form
\begin{equation}
\label{lfieldhub}
H(\phi)=\frac{\Lambda^2}{\sqrt{3M_P^2}}\left(\frac{\phi}{\mu}\right)^{p/2}
\end{equation}
\begin{equation}
\label{lfieldpot}
V(\phi)=\Lambda^4\left(\frac{\phi}{\mu}\right)^{p},
\end{equation}
respectively.

The end of inflation is achieved when $\phi=\phi_e$, whose value can be
determined from (\ref{sol1defp}) and the sign rule for the model $1$ in table \ref{t1}:
\begin{equation}
\label{lfieldphie}
\frac{\phi_e}{M_P}=\frac{p}{\sqrt{2\gamma_e}};
\end{equation}
then, in terms of the expression (\ref{lfieldphie}) the flow parameter $\epsilon$ takes the form
\begin{equation}
\label{lfieldeps}
\epsilon(\phi)=\left[\frac{\sqrt{2\gamma_e}}{p}\right]^{-\beta-2}\left(\frac{\phi}{M_P}\right)^{-\beta-2},
\end{equation}
whereas the two other relevant flow parameters $s$ and $\eta$ are given by
\begin{equation}
\label{lfields}
s(\phi)=\frac{2\beta}{p}\epsilon(\phi),
\end{equation}
\begin{equation}
\label{lfieldeta}
\eta(\phi)=\frac{p-2}{p}\epsilon(\phi),
\end{equation}
where we have used (\ref{defs1}), the first expression of (\ref{eq:flowparams}), plus (\ref{powercs}), (\ref{lfieldhub}) and (\ref{lfieldphie}).
The expression for the number of e-folds is obtained from expressions (\ref{efolds1}), (\ref{efolds2}) and (\ref{sol1alpbet}) for $\alpha\neq0$,
so that
\begin{equation}
\label{lfieldefolds}
N(\phi)=\frac{p}{2\left(\beta+2\right)}\left[\frac{1}{\epsilon(\phi)}-1\right].
\end{equation}

In the analysis performed above we have considered solely models
with $\alpha<0$ and $\beta>0$; the case $\alpha=0$ has been studied in the paper
\cite{tzirakis2008a}, and leads to potentials like
(\ref{lfieldpot}) in the $UV$ limit $s<0$. The case $\beta=0$, $\gamma_e=1$, corresponds to canonical large-field models; in this limit, the expressions
for the flow parameters $\epsilon$ and $\eta$, given by (\ref{lfieldeps}) and (\ref{lfieldeta}), give
\begin{equation}
\label{lfieldceps}
\epsilon(\phi)=\frac{p^2}{2}\frac{M_P^{2}}{\phi^{2}},
\end{equation}
\begin{equation}
\label{lfieldceta}
\eta(\phi)=\frac{p(p-2)}{2}\frac{M_P^{2}}{\phi^{2}},
\end{equation}
which coincides with the results in section III.A of reference \cite{dodelson1997}. Also in this limit, from (\ref{lfieldphie}) we see that inflation
ends when
\begin{equation}
\label{lfieldphiecan}
\frac{\phi_e^{c}}{M_P}=\frac{p}{\sqrt{2}},
\end{equation}
which coincides with the analogous expression found in \cite{dodelson1997}\footnote{Our notation differs
from that adopted in \cite{dodelson1997}; our $M_P$ is related to $m_{pl}$ in that reference by $M_P=m_{pl}/\sqrt{8\pi}$.}. Hence, {\it{all}}
large-field polynomial models with $p>2$
are particular cases of this non-canonical version. Also, if
$\gamma_e\neq 1$, we recover the Spalinski model \cite{spalinski2007b} with a polynomial potential as well. Another particular case of this general
class is {\it{isokinetic inflation}}, proposed in \cite{tzirakis2009a}. For this model, we can show that by setting $\alpha=-p/2-1$ and $\beta=p/2-1$,
we reproduce all the expressions derived in \cite{tzirakis2009a} up to a redefinition of the exponent of the potential\footnote{In isokinetic inflation
the potential has the form $V(\phi)\propto\phi^{2p_{iso}}$ \cite{tzirakis2009a}, whereas in our model we have defined the exponent $p$ (expression
(\ref{sol1defp})) such that $V(\phi)\propto\phi^{p}$. Then $p=2p_{iso}$.}.

Therefore, we have a completely well-defined D-brane inflationary scenario with large-field potentials like (\ref{lfieldpot}), a flow parameter
$\epsilon$ given by (\ref{powerepsilon}) with $\alpha<0$, and a small speed of sound characterized by (\ref{powercs}) with $\beta\geq0$, reproducing
not only the canonical large-field polynomial potentials, but also other models discussed in the literature. For these reasons we will call this class
{\it{non-canonical large-field polynomial models}}.

In particular, as we will see in section \ref{sec:application}, these models predict values for
the scalar spectral index and tensor-to-scalar ratio which agree very well with WMAP5 observations.

\subsection{\label{subsec:smallfield}Small-field polynomial potentials}

Small-field polynomial potentials in canonical inflation arise from a spontaneous symmetry breaking in the presence of a ``false" vacuum in
unstable equilibrium
with nonzero vacuum energy density and a ``physical" vacuum, for which the classical expectation value of the scalar field is nonzero,
$\langle\phi\rangle\neq0$
\cite{Kinney:1995cc}. These models are characterized by an effective symmetry-breaking scale
$\mu\propto\langle\phi\rangle$ such that $\phi\ll\mu\ll M_P$, the field rolls down from an unstable equilibrium at the origin toward
$\mu$; hence, for positive $\phi$ we have always $\dot\phi>0$.

For non-canonical models we can express the small-field limit $\phi\ll\mu$ by choosing $\alpha>0$; also, as we have derived in section
\ref{subsec:sol2}, the condition $\dot\phi>0$ for $\phi>0$ is satisfied when $\sigma=-1$ and $\alpha+\beta+2>0$, which corresponds to the model $5$ in
table \ref{t3}. In this case, the Hubble parameter (\ref{sol2hubpar}) takes the form
\begin{equation}
\label{sfieldeqh}
H(\phi)={\tilde{H}}_e\exp\left[-\frac{K\phi_e}{\alpha+\beta+2}{\bar{\phi}}^{(\alpha+\beta+2)/2}
\right];
\end{equation}
where $K$ and $\phi_e$ are given by (\ref{sol2defKpb}); since $\bar\phi=\phi/\phi_e$ in the small-field limit, and $\alpha+\beta+2>0$,
we can expand expression
(\ref{sfieldeqh}) to first-order in $\bar\phi$, so that
\begin{equation}
\label{sfieldexph}
H(\phi)={\tilde{H}}_e\exp\left[1-\frac{K\phi_e}{\alpha+\beta+2}{\bar{\phi}}^{(\alpha+\beta+2)/2}
\right].
\end{equation}
Since $\beta>-2-\alpha$ and $\alpha>0$, we see that $\beta$ can take either sign; in particular, for $\beta>0$, from (\ref{powercs}) we have the
following relation
\begin{equation}
\label{sfieldgamm1}
\gamma(\phi)=\gamma_e \left(\frac{\phi}{\phi_e}\right)^{-|\beta|} \Longrightarrow \gamma \rightarrow \infty ~~~\mathrm{as}~~~\phi\rightarrow 0,
\end{equation}
or, in terms of the speed of sound,
\begin{eqnarray}
\label{sfieldcs1}
c_s &\rightarrow& 0 ~~~\mathrm{as}~~~\phi\rightarrow 0.
\end{eqnarray}
In the small-field limit, we have always $\phi\ll\phi_e$, so that $\phi\rightarrow 0$ corresponds to early times; then, from (\ref{sfieldcs1})
we conclude that the field propagates with subluminal speed of sound at early times. Also, property (\ref{sfieldgamm1}) implies that
$\gamma/(\gamma+1)\rightarrow 1$, so that
by using this fact and plugging (\ref{sfieldexph}) into (\ref{Vgen}), we find
\begin{equation}
\label{sfieldpot1} V(\phi)\sim 3M_P^2{\tilde{H}_e}^2\left[1-\frac{2}{3}{\bar{\phi}}^{\alpha}
-\frac{2K\phi_e}{\alpha+\beta+2}{\bar{\phi}}^{(\alpha+\beta+2)/2}\right]
\end{equation}
in the slow-roll limit. It is clear that we can derive out of expression (\ref{sfieldpot1}) different sort of potentials, depending on the relations
between the exponents. Let us analyze one of such possible choices; we define first the exponent
\begin{equation}
\label{sfielddefm}
p=\frac{\alpha+\beta+2}{2},
\end{equation}
and we choose $\alpha$ and $\beta$ such that $p$ is always {\it{integer}}. Then, if $\alpha\geq p$, we see that $\alpha\geq\beta+2$, and the potential
(\ref{sfieldpot1}) takes the form
\begin{equation}
\label{sfieldpot2} V(\phi)\sim 3M_P^2{\tilde{H}_e}^2\left[1
-\frac{K\phi_e}{p}{\bar{\phi}}^{p}\right], ~~\alpha\geq\beta+2.
\end{equation}

In the canonical small-field scenario the initial unstable equilibrium state is characterized by the vacuum energy density $\Lambda^4$,
which is the height of the potential at the origin, $\Lambda^4=V(0)$, whereas the effective symmetry-breaking scale is given by \cite{Kinney:1995cc}
\begin{equation}
\label{sfieldwpot1}
\mu=\left.\left[\frac{(m-1)!V(\phi)}{|d^mV/d\phi^m|}\right]^{1/m}\right|_{\phi=0},
\end{equation}
where $m$ is the order of the lowest nonvanishing derivative of the potential at the origin. In the non-canonical
case, the vacuum energy density is given by
\begin{equation}
\label{sfieldvdens}
\Lambda^4=3M_P^2{\tilde{H}}_e^2,
\end{equation}
whereas the effective symmetry-breaking scale
(\ref{sfieldwpot1}) reads
\begin{equation}
\label{sfieldwpot2}
\frac{1}{\mu^{p}}=\sqrt{\frac{2\gamma_e}{M_P^{2}\phi_e^{\alpha+\beta}}};
\end{equation}
then, in terms of these two quantities, the inflationary potential (\ref{sfieldpot1}) becomes, in the small-field limit,
\begin{equation}
\label{sfieldpot}
V(\phi)=\Lambda^4\left[1-\frac{1}{p}\left(\frac{\phi}{\mu}\right)^{p}\right],
\end{equation}
as expected. Then, in the non-canonical case the field also rolls down from an unstable vacuum state whose energy density is given by
(\ref{sfieldvdens}) with very low speed of sound, and evolves toward a minimum characterized by a scale $\mu$ given by (\ref{sfieldwpot2})
for $\alpha\geq p\geq 2$, and such behavior is exactly the same as the canonical case.

From (\ref{sfieldwpot2}) we see that inflation ends when
\begin{equation}
\label{sfieldphie}
\frac{\phi_e}{\mu}=\left[\frac{\mu}{M_P}\sqrt{2\gamma_e}\right]^{1/(p-1)},
\end{equation}
so that the flow parameters are given by
\begin{equation}
\label{sfieldeps}
\epsilon(\phi)=\left[\frac{M_P}{\mu\sqrt{2\gamma_e}}\right]^{(p-1)/\alpha}\left(\frac{\phi}{\mu}\right)^{\alpha},
\end{equation}
\begin{equation}
\label{sfields}
s(\phi)=-\beta\sqrt{\frac{2M_P^{2}}{\gamma_e\phi_e^{2}}},
\end{equation}
\begin{equation}
\label{sfieldeta}
\eta(\phi)=(\alpha+\beta)\sqrt{\frac{M_P^{2}}{2\gamma_e\phi_e^{2}}}{\bar{\phi}}^{(\alpha-\beta-2)/2}+\epsilon(\phi),
\end{equation}
where we have used (\ref{defs1}), the first expression of (\ref{eq:flowparams}), plus (\ref{powerepsilon}), (\ref{powercs}), (\ref{sfieldeqh})
and (\ref{sfieldphie}).

In particular, in the canonical limit $\beta=0$, $\gamma_e=1$ we have $\alpha\geq 2$, so that for {\it{even}} values of $\alpha$ all the potentials with
$p\geq 2$ are reproduced. In this limit, from (\ref{sfielddefm}) we see that $\alpha=2(p-1)$; then, from (\ref{sfieldeps})
the flow parameter $\epsilon$ assumes the form
\begin{equation}
\label{sfieldceps}
\epsilon(\phi)=\frac{M_P}{\mu\sqrt{2}}\left(\frac{\phi}{\mu}\right)^{2(p-1)},
\end{equation}
whereas from (\ref{sfieldphie}) we see that inflation ends at
\begin{equation}
\label{sfieldcphie}
\frac{\phi_e}{\mu}=\left[\frac{\mu}{M_P}\sqrt{2}\right]^{1/(p-1)}.
\end{equation}
Expressions (\ref{sfieldceps}) and (\ref{sfieldcphie}) agree with the corresponding expressions (3.25) and (3.26) in \cite{Kinney:1995cc} derived
for canonical small-field potentials.

Therefore, in the slow-roll limit {\it{all}} canonical small-field polynomial models with $p \geq 2$ are particular solutions to the non-canonical model
described in this
section when $\beta=0$, $\gamma_e=1$ and $\alpha$ even; hence, we have
again a well-defined D-brane inflationary scenario with a small-field potential like (\ref{sfieldpot}), a flow parameter
$\epsilon$ given by (\ref{powerepsilon}) with $\alpha>0$, and a small speed of sound characterized by (\ref{powercs}) with $\beta\leq0$, reproducing
all the canonical small-field polynomial potentials when $\beta=0$. For these reasons we will call this class
{\it{non-canonical small-field polynomial models}}.

\subsection{\label{subsec:hybpot}Hybrid potentials}

In the last two sections we have discussed the small-field models characterized by $\dot\phi>0$ for positive $\phi$, given by the model $5$ in table
\ref{t3}. Let us now examine a similar model, with $\alpha+\beta+2>0$ but with $\dot\phi<0$ for positive $\phi$. In this case, $\sigma=+1$,
(model $6$ in table \ref{t3}); then, the Hubble parameter (\ref{sol2hubpar}) takes the form
\begin{equation}
\label{hybhub1}
H(\phi)={\tilde{H}}_e\exp\left[\frac{K\phi_e}{\alpha+\beta+2}{\bar{\phi}}^{(\alpha+\beta+2)/2}
\right],
\end{equation}
where the constant $K$ and variable $\bar\phi$ are given by the definitions (\ref{sol2defKpb}). Since $\bar\phi\ll 0$ and
$p>0$, where $p$ is given by (\ref{sfielddefm}), we expand expression
(\ref{hybhub1}) to first-order in $\bar\phi$,
\begin{equation}
\label{hybhub}
H(\phi)={\tilde{H}}_e\exp\left[1+\frac{K\phi_e}{2p}{\bar{\phi}}^{p}
\right].
\end{equation}
The analysis leading to the sign of $\beta$ is identical to that made in section \ref{subsec:smallfield} since, as in that case, $\beta>-2-\alpha$ and
$\alpha>0$; then, using the same arguments we find that the field rolls down the potential with a subluminal speed of sound. Also, since
$\gamma/(\gamma+1)\rightarrow 1$ at early times, we have, plugging (\ref{hybhub}) into (\ref{Vgen}), that
\begin{equation}
\label{hybpot1}
V(\phi)\sim 3M_P^2{\tilde{H}_e}^2\left[1-\frac{2}{3}{\bar{\phi}}^{\alpha}
+\frac{K\phi_e}{p}{\bar{\phi}}^{p}\right]
\end{equation}
in the slow-roll limit. As in the model derived in section \ref{subsec:smallfield}, we choose $\alpha$ and $\beta$ such that $p$ is always integer;
then, for $\alpha>p$, the potential
(\ref{hybpot1}) takes the form
\begin{equation}
\label{hybpot2} V(\phi)\sim 3M_P^2{\tilde{H}_e}^2\left[1
+\frac{K\phi_e}{p}{\bar{\phi}}^{p}\right].
\end{equation}
In this case, the minimum of the potential is at the origin, as in the small-field polynomial case, but now $V(\phi)>V(0)$ around $\phi=0$. Therefore,
the field rolls toward the minimum with nonzero vacuum energy, $\Lambda^4=V(0)$. This is exactly the behavior of the canonical hybrid potentials
\cite{Linde:1991km,Linde:1993cn}. Then, we may write the potential (\ref{hybhub1}) as
\begin{equation}
\label{hybpot}
V(\phi)\sim \Lambda^4\left[1+\left(\frac{\phi}{\mu}\right)^{p}\right],
\end{equation}
where we have defined
\begin{equation}
\label{hybmu}
\frac{1}{\mu^{p}}=\frac{1}{p}\sqrt{\frac{2\gamma_e}{M_P^{2}\phi_e^{2(p-1)}}}.
\end{equation}

\subsection{\label{subsec:exppot}Exponential potentials}

Along with the models discussed above, there is a fourth type whose Hubble parameter and potential are exponentials \cite{dodelson1997}. Since the
general expression for the Hubble parameter derived in section \ref{subsec:sol2} is of a exponential form, we focus on its large-field limit solution,
given by the model $4$ in table \ref{t2}. In this case $\dot\phi<0$ for positive $\phi$, so that $\sigma=+1$. The expression for the Hubble parameter
(\ref{sol2hubpar}) for $\alpha<0$, is given by
\begin{equation}
\label{exppothub1}
H(\phi)={\tilde{H}_e}\exp\left[\sqrt{\frac{1}{2p}\left(\frac{\phi}{M_P}\right)^{\alpha+\beta+2}}\right],
\end{equation}
where we have defined
\begin{equation}
\label{exppotdefp}
p=\frac{\left(\alpha+\beta+2\right)^2}{4\gamma_e}\left(\frac{\phi_e}{M_P}\right)^{\alpha+\beta}.
\end{equation}

Since $\alpha+\beta+2>0$, the exponent of the speed of sound is restricted to the values $\beta>-\alpha-2$; then, for $\beta>0$, we have that
$\gamma \rightarrow \infty$ as $\phi\rightarrow \infty$, and then $c_s \rightarrow 0$ at early times since $\phi$ is in the large-field limit. Hence,
the field propagates with a subluminal speed of sound at early times, and $\gamma/(\gamma+1)\rightarrow 1$. Using this fact and substituting
(\ref{exppothub1}) into (\ref{Vgen}) we find
\begin{equation}
\label{exppotpot1} V(\phi)\sim 3M_P^2{\tilde{H}}_e^2\exp\left[\sqrt{\frac{2}{p}\left(\frac{\phi}{M_P}\right)^{\alpha+\beta+2}}\right].
\end{equation}
Then, the field rolls down the potential toward the minimum at origin, characterized by a nonzero vacuum energy $V(0)=\Lambda^4$
with a subluminal speed of sound. This is similar to the behavior of exponential potentials in canonical models, except for the fact that $c_s=1$.
Since $\Lambda^4=3M_P{\tilde{H}_e}^2$, the final form of the non-canonical potential (\ref{exppotpot1}) is
\begin{equation}
\label{exppotpot} V(\phi)\sim \Lambda^4\exp\left[\sqrt{\frac{2}{p}\left(\frac{\phi}{M_P}\right)^{\alpha+\beta+2}}\right].
\end{equation}

Before we study the non-canonical limit of the potential (\ref{exppotpot}), let us have a look first at the flow-parameter $\epsilon$.
We have used the parametrization associated with $\alpha\neq0$, given by (\ref{powerepsilon}), but we can make it general as follows: substituting
and (\ref{powercs}) and (\ref{exppotdefp}) into (\ref{powerepsilon}), we find
\begin{equation}
\label{exppoteps}
\epsilon(\phi)=\frac{\left(\alpha+\beta+2\right)^2}{4 p\gamma(\phi)}\left(\frac{\phi}{M_P}\right)^{\alpha+\beta}.
\end{equation}
which holds even when $\alpha=0$, for $\epsilon(\phi)=\epsilon=const.$ in that case. The other two flow parameters $s$ and
$\eta$ are given respectively by
\begin{equation}
\label{exppots}
s(\phi)=\beta\left(\frac{\phi}{M_P}\right)^{-1}\sqrt{\frac{2\epsilon(\phi)}{\gamma(\phi)}},
\end{equation}
\begin{equation}
\label{exppoteta}
\eta(\phi)=\frac{\alpha+\beta}{\sqrt{2}}\left(\frac{\phi}{M_P}\right)^{-1}\sqrt{\frac{\epsilon(\phi)}{\gamma(\phi)}}+\epsilon(\phi),
\end{equation}
where we have used (\ref{defs1}), the first expression of (\ref{eq:flowparams}), plus (\ref{powercs}) and (\ref{exppothub1}).

Then, with the parametrization defined by (\ref{exppoteps}), we see that in the canonical case $\alpha=\beta=0$, $\gamma_e=1$, expressions
(\ref{exppoteps}) and (\ref{exppoteta}) give
\begin{equation}
\label{exppotcl}
\epsilon(\phi)=\eta(\phi)=\frac{1}{p},
\end{equation}
which matches the results derived for the canonical case as shown in \cite{dodelson1997}. Expression (\ref{exppotcl}) shows that we have to restrict the
values of (\ref{exppotdefp}) to be $p>1$, so that we get $\epsilon\leq 1$ in the canonical limit.

Therefore, we have a completely well-defined D-brane
inflationary scenario with exponential potentials like (\ref{exppotpot}), a flow parameter
$\epsilon$ given by (\ref{exppoteps}) with $\alpha\leq0$, and a small speed of sound characterized by (\ref{powercs}) with $\beta\geq0$, which
reproduces the corresponding canonical model. We will call this class
{\it{non-canonical exponential models}}.


The four distinct non-canonical classes obtained so far are summarized in Table \ref{t4} below.

\begin{table*}
\begin{center}
\caption{A summary of the distinct models discussed in this work.}\label{t4}
\begin{tabular}{c|c|c|c|c}
\hline \hline
 & & & &  \\
model &
$\alpha$ &
$\beta$ &
$p$ & $V(\phi)$\\
 & & & & \\
\hline \hline & & & & \\
Large-field &
$\alpha=-\beta-2$ &
$\beta\geq 0$ &
$\phi_e\sqrt{\frac{2\gamma_e}{M_P^{2}}}$$~~~$\footnote{Corresponds to model $1$ in table \ref{t1}.} &
$\Lambda^4\left(\frac{\phi}{\mu}\right)^{p}$ \\
 & & & & \\
Small-field &
$\alpha\geq p$ &
$\beta\leq 0$ &
$\frac{\alpha+\beta+2}{2}$ $~~$\footnote{$p$ is integer. Corresponds to model $5$ in table \ref{t3}.}&
$\Lambda^4\left[1-\frac{1}{p}\left(\frac{\phi}{\mu}\right)^{p}\right]$ \\
 & & & & \\
Hybrid &
$\alpha\geq p$ &
$\beta\leq 0$ &
$\frac{\alpha+\beta+2}{2}$ $~~~$\footnote{$p$ is integer. Corresponds to model $6$ in table \ref{t3}.}&
$\Lambda^4\left[1+\left(\frac{\phi}{\mu}\right)^{p}\right]$ \\
 & & & & \\
Exponential &
$\alpha\leq 0$ &
$\beta\geq 0$ &
$\frac{\left(\alpha+\beta+2\right)^2}{4\gamma_e}\left(\frac{\phi_e}{M_P}\right)^{\alpha+\beta}$ $~~~$\footnote{$p>1$. Corresponds to model
$4$ in table \ref{t2}.}&
$\Lambda^4\exp\left[\sqrt{\frac{2}{p}\left(\frac{\phi}{M_P}\right)^{\alpha+\beta+2}}\right]$ \\
 & & & & \\ \hline \hline
\end{tabular}
\end{center}
\end{table*}

\section{\label{sec:application}An application of non-canonical large-field polynomial models}

In this section we study some applications of the large-field models derived in section \ref{subsec:largefield}. We choose this class of non-canonical
potentials because the expressions for the scalar spectral index, the tensor/scalar ratio and the level of non-gaussianity are particularly simple,
depending on two
parameters solely, $p$ and $\beta$. Let us first derive an expression for the flow parameter $\epsilon$ in terms of $N$. From (\ref{lfieldeps})
and (\ref{lfieldefolds}) we find
\begin{equation}
\label{aplfieldepsN}
\epsilon(N)=\frac{p}{p+2\left(\beta+2\right)N}.
\end{equation}
The other two flow parameters $s$ and $\eta$ are given by
\begin{equation}
\label{aplfields}
s(N)=\frac{2\beta}{p+2\left(\beta+2\right)N},
\end{equation}
\begin{equation}
\label{aplfieldeta}
\eta(N)=\frac{p-2}{p+2\left(\beta+2\right)N};
\end{equation}
where we have used (\ref{defs1}), the first expression of (\ref{eq:flowparams}), plus (\ref{powercs}), (\ref{lfieldhub}) and (\ref{aplfieldepsN}).
Inserting (\ref{aplfieldepsN}), (\ref{aplfields}) and (\ref{aplfieldeta}) into (\ref{scalarspecindex}), we find, in the slow-roll limit,
\begin{equation}
\label{aplfieldns}
n_s=1-\frac{2(p+2\beta+2)}{p+2\left(\beta+2\right)N}.
\end{equation}

The expression for the speed of sound in terms of $N$ can be calculated in the same way: we use
(\ref{powerepsilon}), (\ref{powercs}) and
(\ref{lfieldefolds}), so that
\begin{equation}
\label{aplfieldcsN}
c_s(N)=\frac{1}{\gamma_e}\left[\frac{p}{p+2\left(\beta+2\right)N}\right]^{\beta/(\beta+2)}.
\end{equation}
Through the use of (\ref{stratio}), (\ref{aplfieldepsN}) and (\ref{aplfieldcsN}) we can derive a general expression for the tensor/scalar ratio,
which is given by
\begin{equation}
\label{aplfieldr}
r(N)=\frac{16}{\gamma_e}\left[\frac{p}{p+2\left(\beta+2\right)N}\right]^{2(\beta+1)/(\beta+2)}.
\end{equation}

The expression for the level of non-gaussianity $f_{NL}$ is given by \cite{peiris2007}
\begin{equation}
\label{aplfieldfnl}
f_{NL}=-\frac{35}{108}\left(\frac{1}{c_s^2}-1\right),
\end{equation}
which can be easily evaluated by using expression (\ref{aplfieldcsN}).

\begin{figure*}
\centering
\includegraphics[scale=0.50]{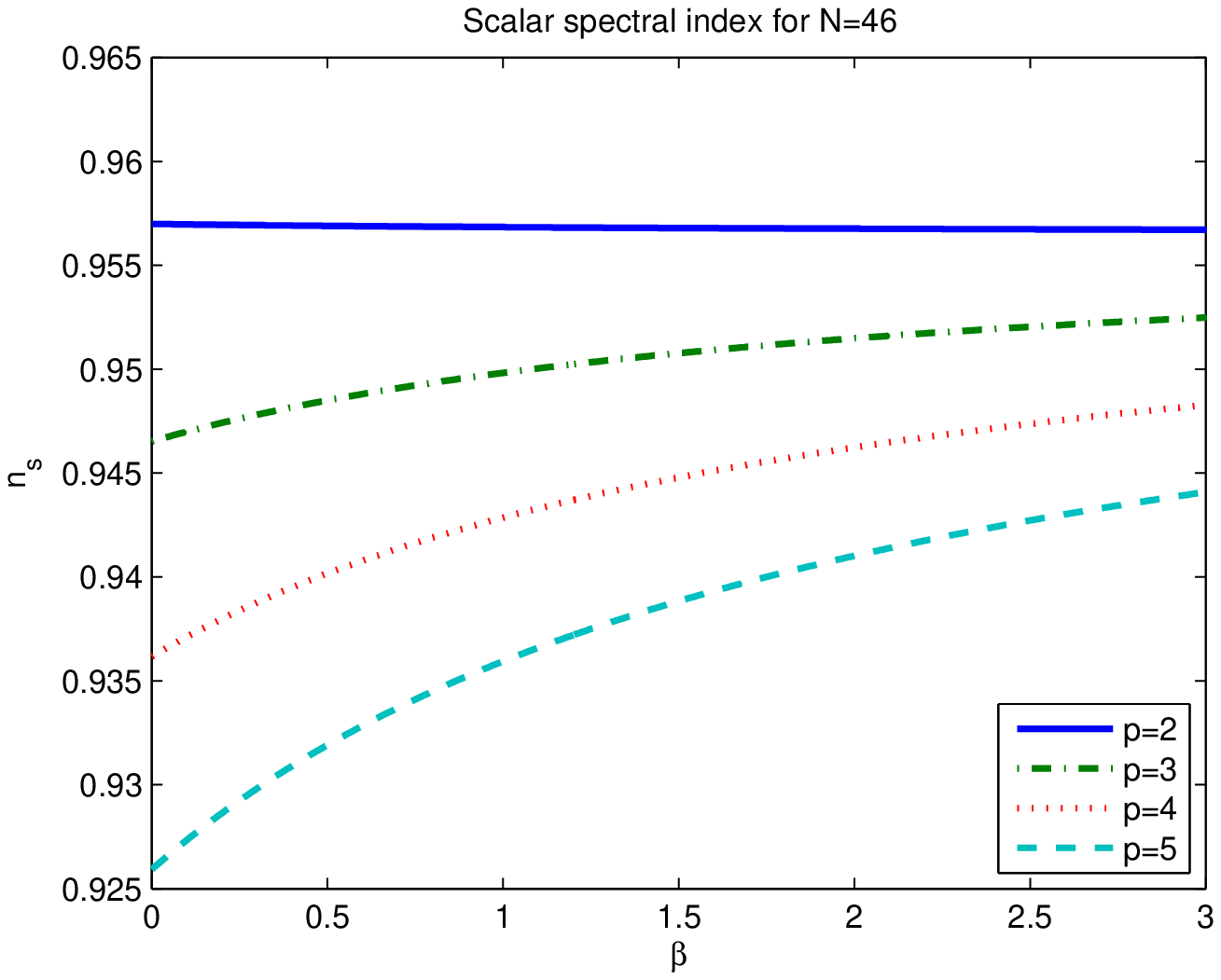}
\includegraphics[scale=0.50]{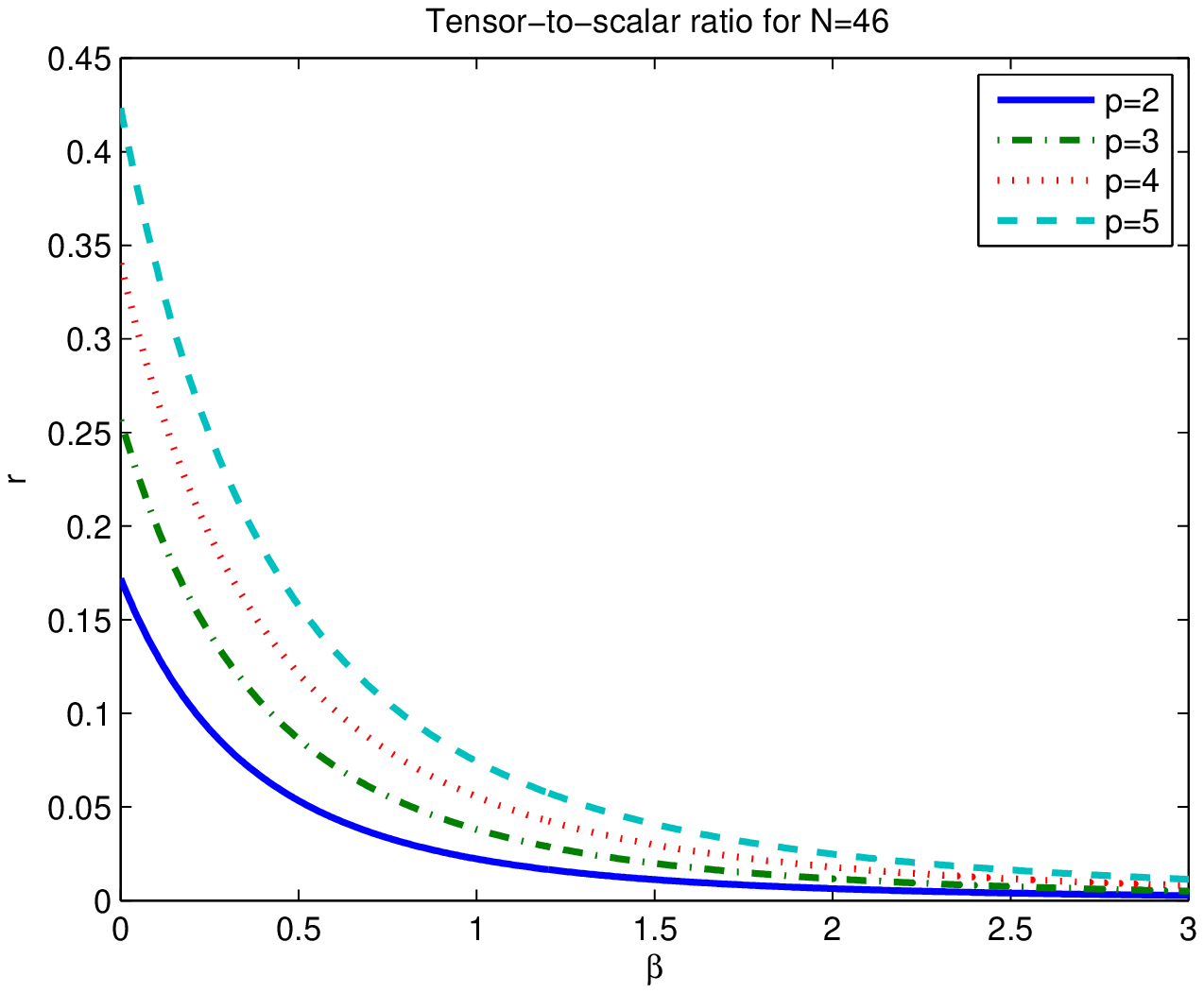}
\includegraphics[scale=0.50]{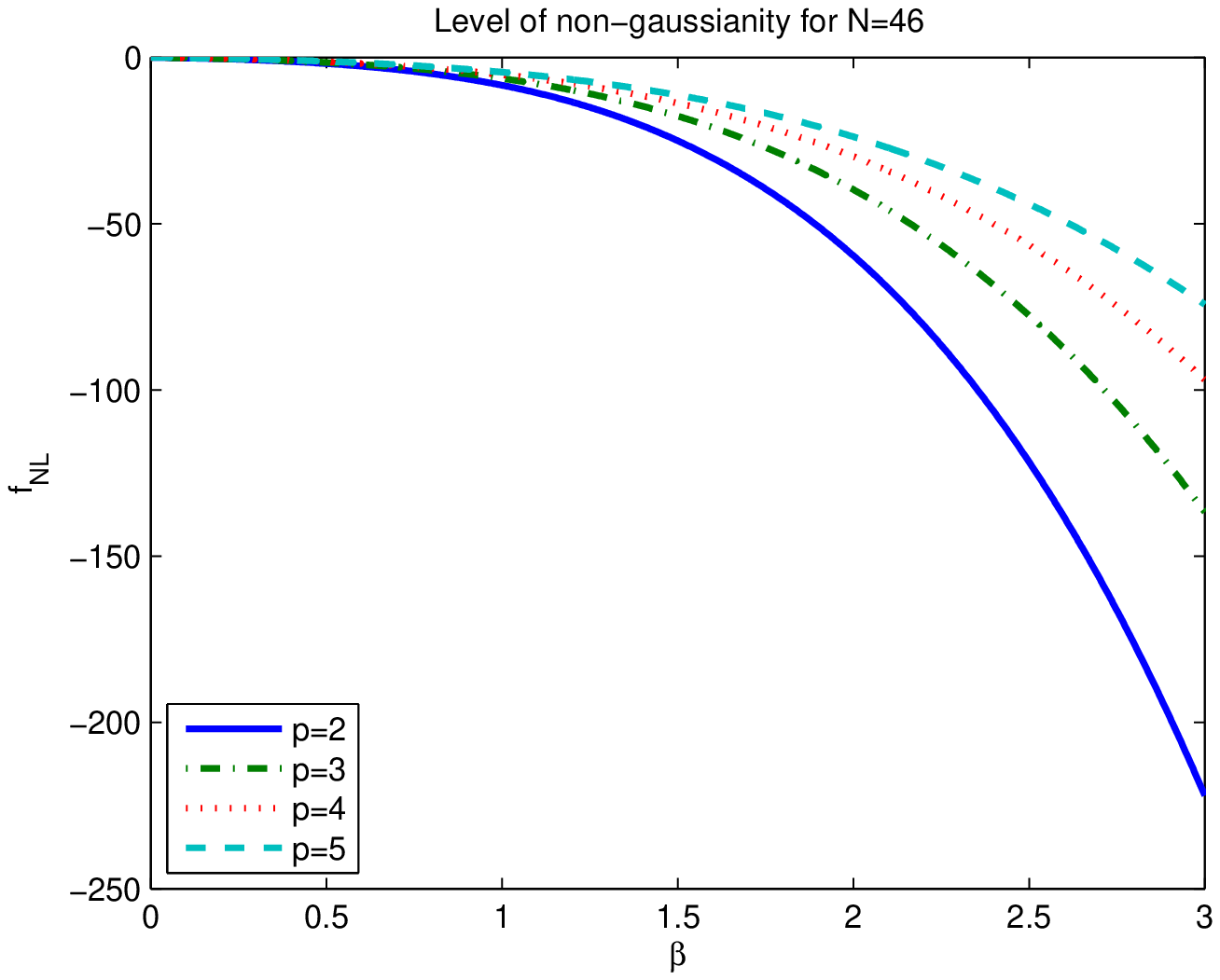}
\includegraphics[scale=0.50]{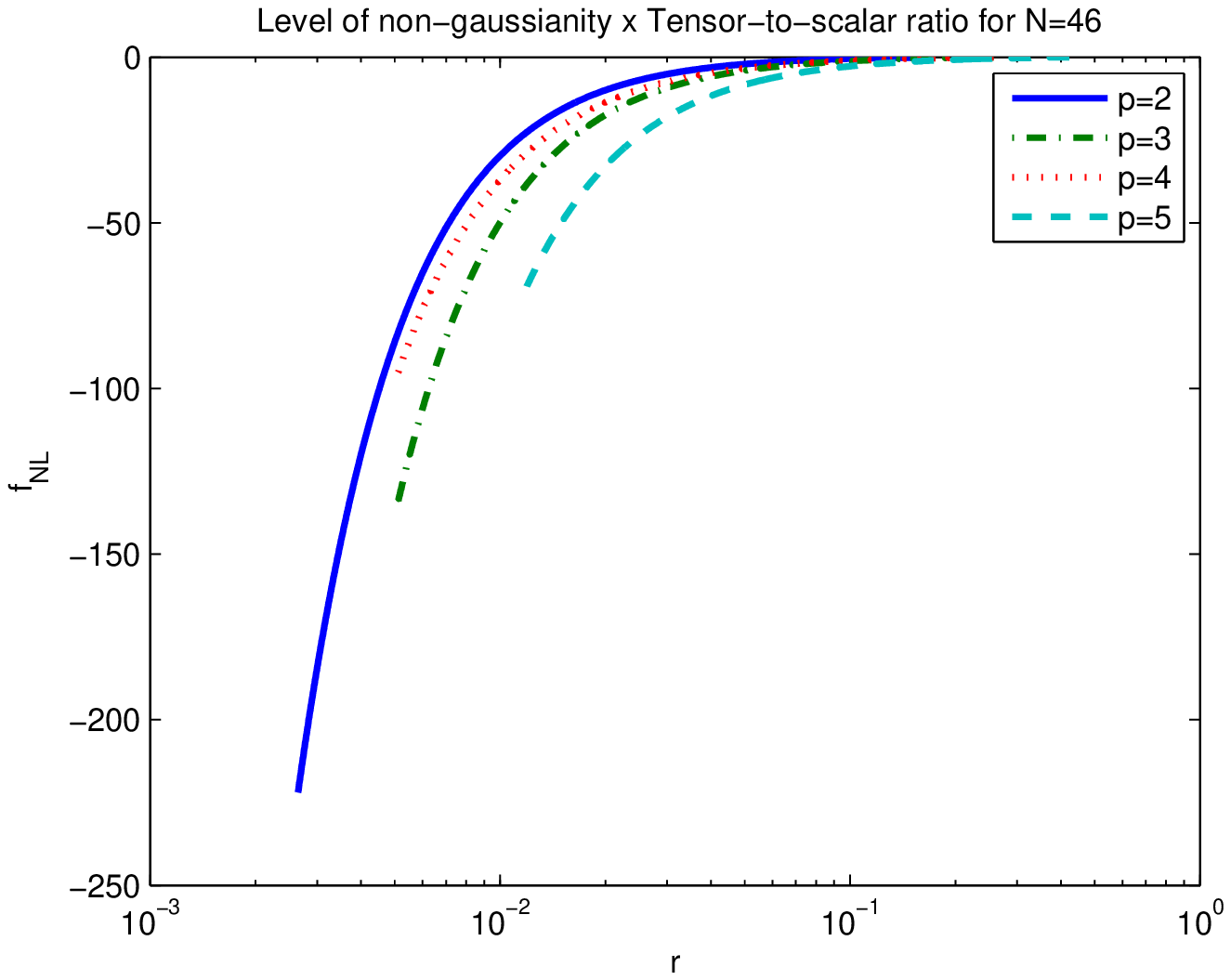}
\caption{The observables $n_s$ (top left), $r$ (top right) and $f_{NL}$ (bottom left) as a function of the exponent of the speed of sound $\beta$
for each value of $p$ ($V(\phi)\propto\phi^p$) for $N=46$. On bottom right is depicted the behavior of $f_{NL}$ compared to $r$. \label{fig:nsrfnlN46}}
\end{figure*}

\begin{figure*}
\centering
\includegraphics[scale=0.50]{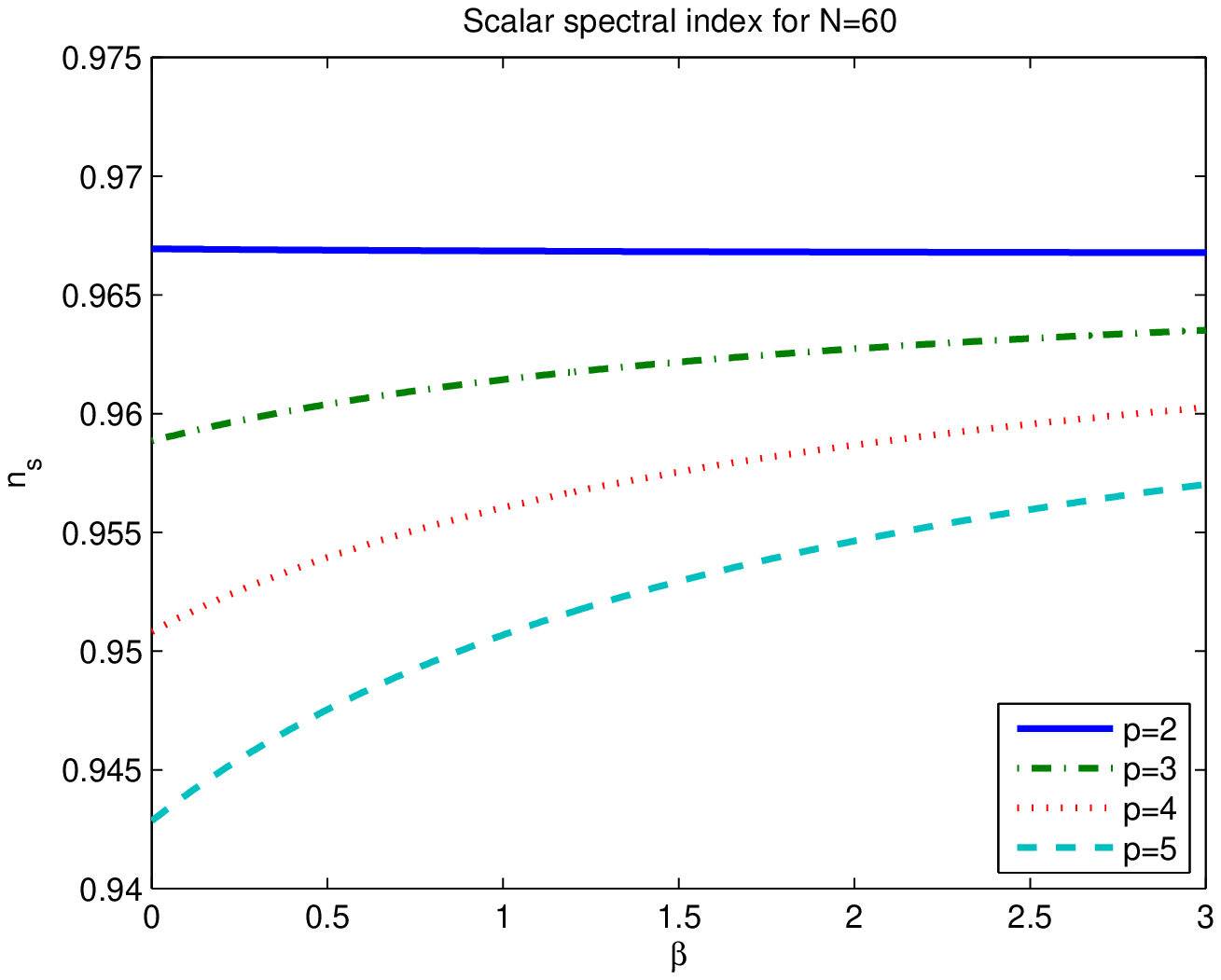}
\includegraphics[scale=0.50]{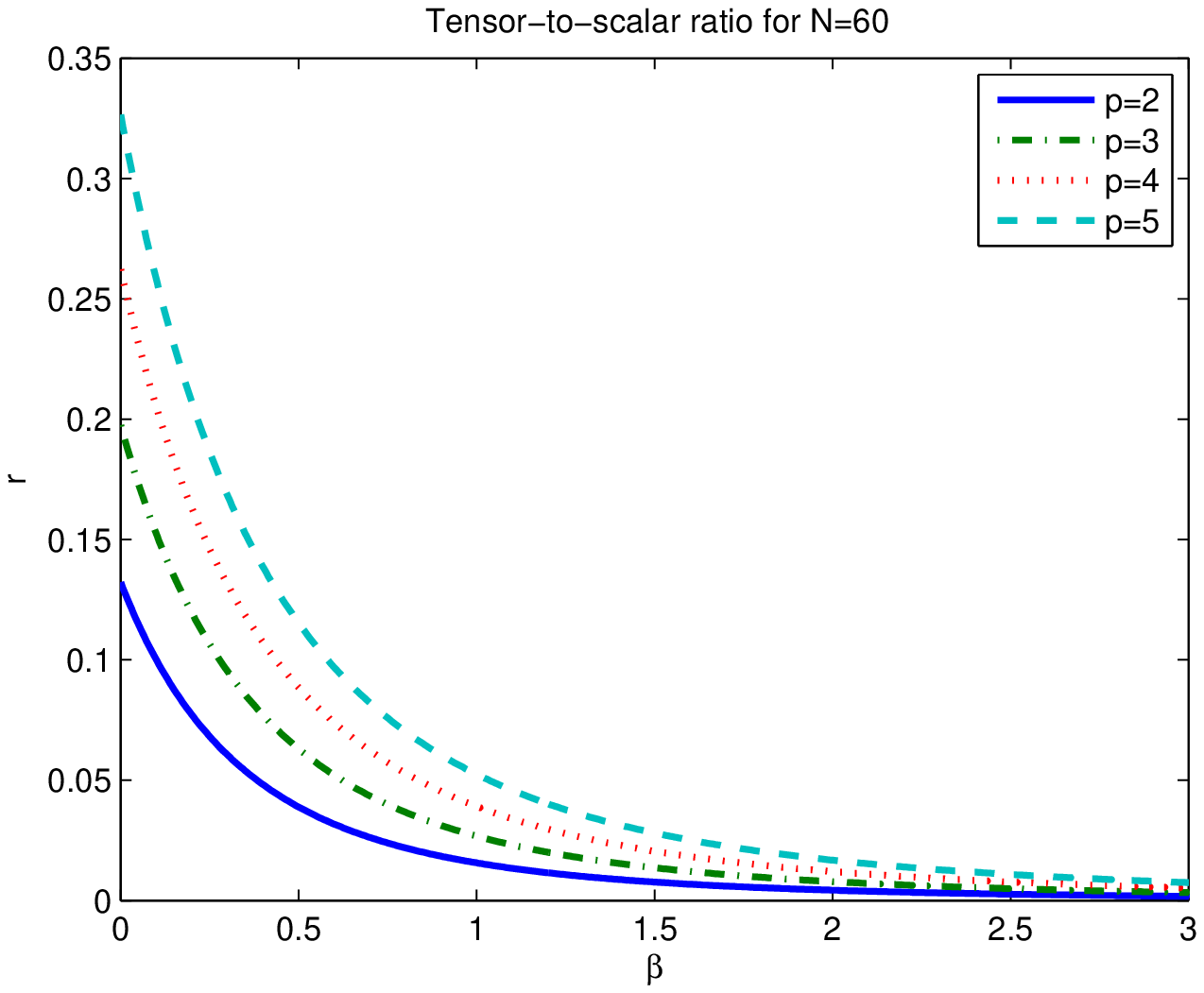}
\includegraphics[scale=0.50]{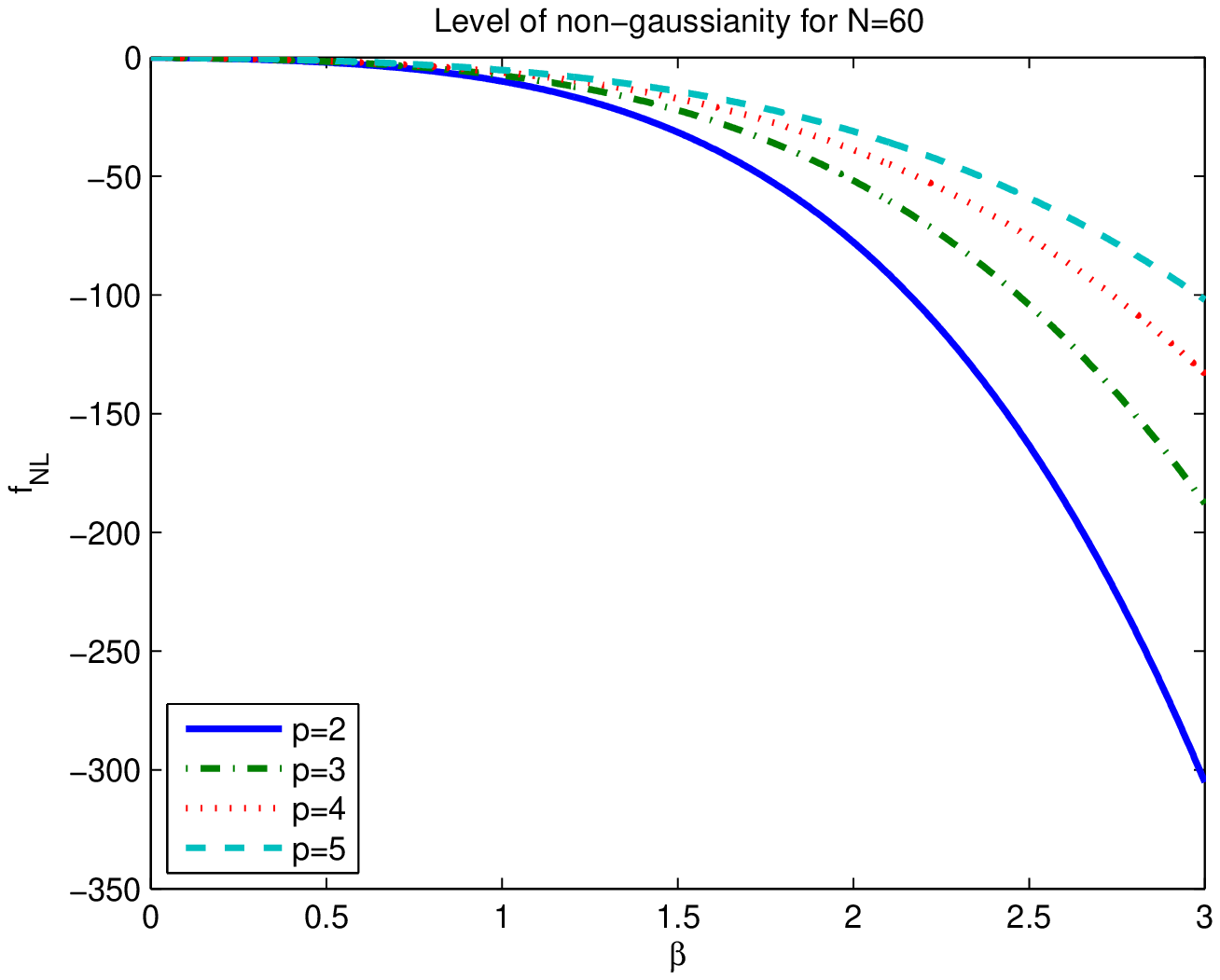}
\includegraphics[scale=0.50]{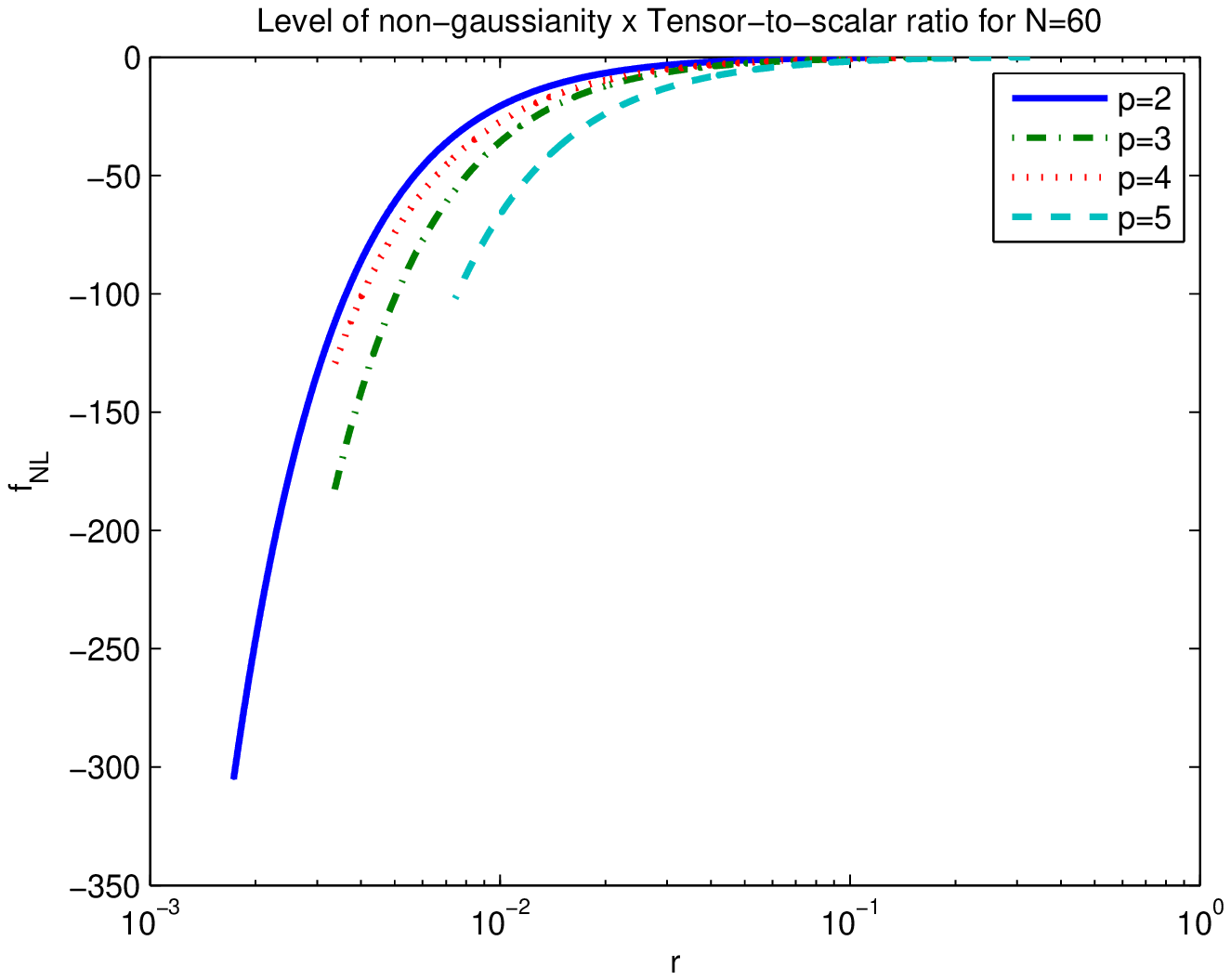}
\caption{The observables $n_s$ (top left), $r$ (top right) and $f_{NL}$ (bottom left) as a function of the exponent of the speed of sound $\beta$
for each value of $p$ ($V(\phi)\propto\phi^p$) for $N=60$. On bottom right is depicted the behavior of $f_{NL}$ compared to $r$. \label{fig:nsrfnlN60}}
\end{figure*}

\begin{figure*}
\centering
\includegraphics[width=1\textwidth]{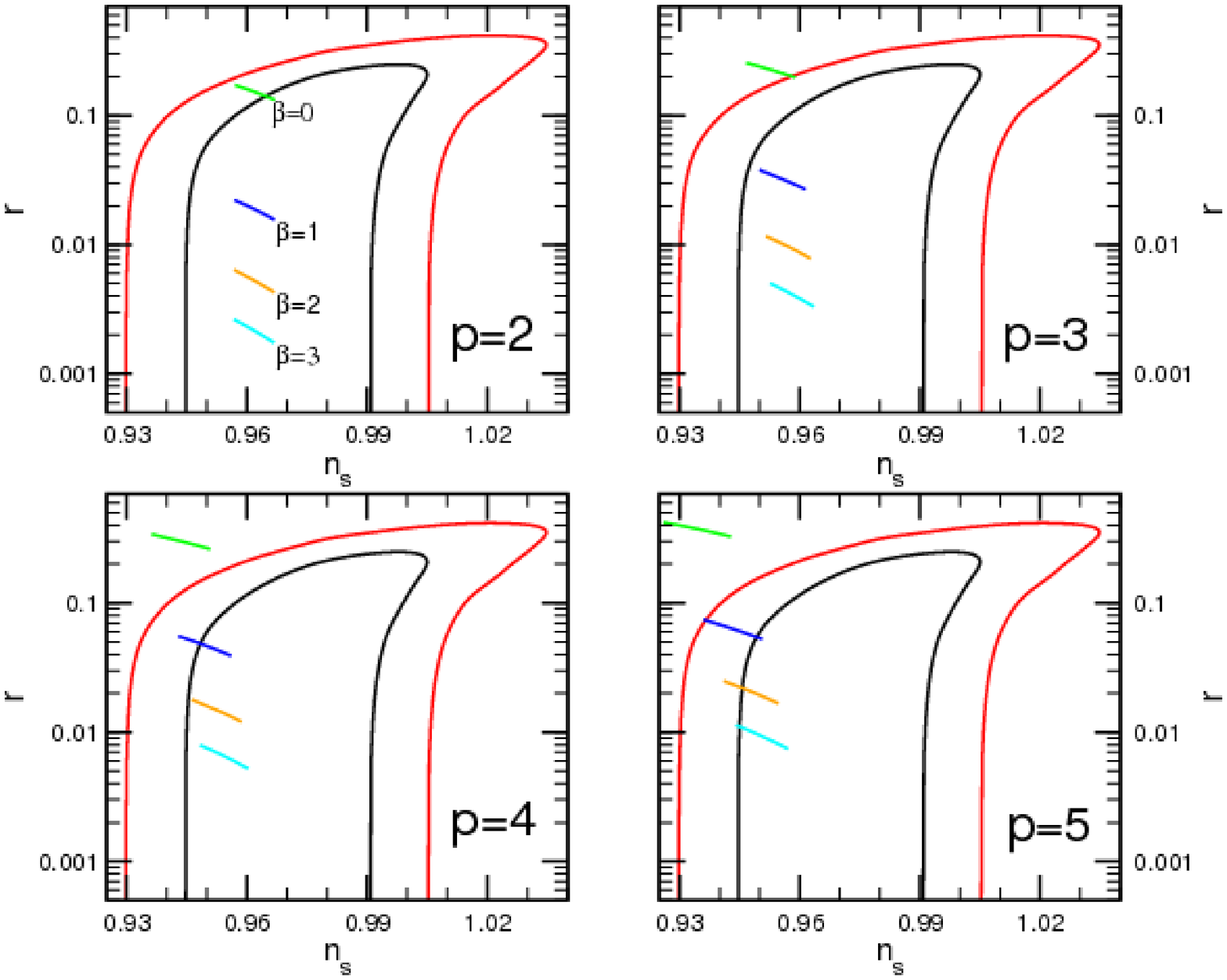}
\caption{68 \% (black) and 95 \% C.L. (red) on the $n_{s}$ and
$r$ parameter space for WMAP5 alone. In each panel we plot the values of $n_s$ and $r$ for a specific potential $V(\phi)\propto\phi^p$ according to
the exponent $\beta$ of the
speed of sound. Green lines correspond to $\beta=0$ (canonical limit), blue lines to $\beta=1$, orange lines to $\beta=2$ and light blue lines to
$\beta=3$. The left (right) extremity of each line correspond to the case where a mode
crossed the sound horizon $46$ ($60$) e-folds before the end of
inflation.} \label{fig:nrplot}
\end{figure*}

Therefore, the tensor/scalar ratio will have a power-law dependence as well, with exponent $2(\beta+1)/(\beta+2)$, which means that, for a given
value of $p$, a larger $\beta$ corresponds to a smaller $r$. Since $\beta\geq 0$ for non-canonical large-field models, we have, from (\ref{gammafac})
and (\ref{powercs}), that $c_s\propto\phi^{-\beta}$; then, fields rolling with {\it{slower}} speed of sound would produce {\it{lower}}
tensor/scalar ratios. However, from (\ref{aplfieldfnl}), we see that $f_{NL}$ depends on $c_s^{-2}$, and then a low speed of sound would produce
a {\it{larger}} level of non-gaussianity; then, for large-field models low-$r$ tensor modes are strongly correlated with the amplitude of
non-gaussianity, as has been discussed in the reference \cite{tzirakis2009a} for isokinetic inflation. Then, the suppression of tensor modes by
a large amount of non-gaussianity is a feature shared by all non-canonical models with large-field polynomial potentials.

Let us next make some predictions on the values of $n_s$, $r$ and $f_{NL}$ through expressions (\ref{aplfieldns}), (\ref{aplfieldr}) and
(\ref{aplfieldfnl}) respectively, when the
modes cross the horizon
$46$ or $60$ e-folds before the end of inflation. As we have
discussed in section \ref{subsec:gensetting}, we have set $\gamma_e=1$, which characterizes the end of inflation; the results are depicted in figures
\ref{fig:nsrfnlN46} and \ref{fig:nsrfnlN60}. In both figures, the top left plots refer to the variation of the scalar index in terms
of $\beta$ for each value of $p$. The top right plots in Figs. \ref{fig:nsrfnlN46} and \ref{fig:nsrfnlN60} show the corresponding tensor/scalar ratio. In these plots we
see that for larger values of $p$ and small $\beta$ the modes have large values of $r$ (the observable lower bound is $r<0.22$); then, as $\beta$
increases, the speed of sound gets lower and, in consequence, the tensor/scalar ratio as well. However, as shown in (\ref{aplfieldfnl}), a field
rolling very slowly produces a large amount of non-gaussianity, as can be seen in the bottom left plots of figures \ref{fig:nsrfnlN46} and
\ref{fig:nsrfnlN60}. Therefore, as was first discussed in the particular case of isokinetic inflation \cite{tzirakis2009a}, the production of large
non-gaussianity is strictly correlated with low tensor amplitudes, and this is a feature common to all large-field polynomial potentials.
This behavior is shown in the bottom right plots of figures \ref{fig:nsrfnlN46} and \ref{fig:nsrfnlN60}.

We next compare the results obtained with the current WMAP5 data \cite {komatsu2009}, \cite{kinney2008}. The results are depicted in Fig.
\ref{fig:nrplot} for different values of $p$. Straight lines indicate the different values of $\beta$, with the left (right) extremity indicating the
value of $(n_s,r)$ evaluated at $N=46$ ($N=60$). As shown in \cite{kinney2008}, all canonical models with $p>2$ are ruled out by WMAP5 data alone;
however, in the non-canonical case, figure \ref{fig:nrplot} shows that the models with $p\leq5$ are also consistent with the observable data. A field
evolving with slow-varying speed of sound produces low-amplitude tensors, then pushing the values $(n_s,r)$ inwards the observable region. However,
a large amount of non-gaussianity is produced, which is a distinct signature of non-canonical large-field polynomial models and
can be a powerful observable to discriminate among inflationary models.

\section{\label{sec:conclusions}Conclusions}

In this paper we propose a general DBI model characterized by a power-law flow-parameter power-law flow parameter
$\epsilon(\phi)\propto\phi^{\alpha}$ and speed of sound $c_s(\phi)\propto\phi^{\beta}$, where $\alpha$ and $\beta$ are constants. We show that this
general model has distinct classes of solutions depending on the relation between $\alpha$ and $\beta$, and on the time evolution of the inflaton field.
These classes of solutions are summarized in tables \ref{t1}, \ref{t2} and \ref{t3}. In particular, we show that in the slow-roll limit the four
well-known canonical potentials arise naturally in this general DBI model, having similar properties to their canonical counterparts, except that the speed of sound in general varies with time. We also show that this general DBI model encompasses not only all the canonical
models with the mentioned potentials, but other D-brane scenarios as well: the DBI model with constant speed of sound \cite{spalinski2007b}, with
constant flow parameters \cite{tzirakis2008a}, and isokinetic inflation \cite{tzirakis2009a}. The four non-canonical models are summarized in table
\ref{t4}.

We also derive the expressions for the spectral index, tensor/scalar ratio and the amplitude of non-gaussianity for large-field potentials
in the slow-roll limit. We show that a low speed of sound suppresses the tensor/scalar ratio $r$ and produces a large amount of
non-gaussianity, a feature already explored in the case of isokinetic inflation, and shown to be a general property of all large-field DBI models with
polynomial potentials.
Unlike canonical inflation, where all polynomial models with $p>2$ are ruled out, the suppression of tensor modes in the non-canonical version allows
a larger class of polynomial potentials to lie within the observable range; also, the production of large amount of non-gaussianity is a distinct
signature of these DBI large-field models, which can be a powerful observable to discriminate among inflationary models.

\begin{acknowledgements}

DB thanks Brazilian agency CAPES for financial support. This research is supported in part by the National Science Foundation under grant NSF-PHY-0757693.

\end{acknowledgements}

\newpage


\end{document}